\documentclass[twocolumn, traditabstract]{aa}
\usepackage{color}
\usepackage{soul}
\usepackage{graphicx}
\usepackage{txfonts}
\usepackage{lscape}
\usepackage{longtable}
\usepackage{threeparttable}
\usepackage{setspace}
\usepackage{caption}
\sethlcolor{yellow}

\begin{document}
   \title{Black hole masses and starbursts in X-shaped radio sources}

   \author{M. Mezcua\inst{1}
	\thanks{Email:mmezcua@mpifr.de \newline Member of the International Max Planck Research School (IMPRS)
	for Astronomy and Astrophysics at the Universities of Bonn and
	Cologne.}
   \and A.P. Lobanov\inst{1}
   \and V.H. Chavushyan\inst{2}
   \and J. Le\'on-Tavares\inst{1,2,3}}
   \institute{Max Planck Institute for Radio Astronomy,
              Auf dem H\"ugel 69, 53121 Bonn
   \and Instituto Nacional de Astrof\'isica, \'Optica y Electr\'onica, Apdo. Postal 51, 72000 Puebla, M\'exico
   \and Aalto University Mets\"ahovi Radio Observatory, Mets\"ahovintie 114, FIN-02540 Kylm\"al\"a, Finland}

  \abstract{It has been suggested that the X-shaped
morphology observed in some radio sources can reflect either a recent merger of
two supermassive black holes (SMBHs) or the presence of a second
active black hole in the galactic nucleus.  These scenarios are tested by
studying the relationship between the black hole mass, radio and optical
luminosity, starburst history, and dynamic age of radio lobes in a
sample of 29 X-shaped radio galaxies drawn from a list of 100 X-shaped
radio source candidates identified from the FIRST
survey. The same relationships are also studied in a control sample
consisting of 36 radio-loud active nuclei with similar redshifts and
optical and radio luminosities. The X-shaped objects are found to have
statistically higher black hole masses and older starburst activity
compared with the objects from the control sample. Implications
of these findings are discussed for the black hole merger
scenario and for the potential presence of active secondary black
holes in post-merger galaxies.}

 \keywords{Galaxies: kinematics and dynamics -- Galaxies: formation -- Galaxies: nuclei -- Black hole physics}

\maketitle

\section{Introduction}

X-shaped (or `winged') radio galaxies are a class of extragalactic
radio sources with two low-surface-brightness radio lobes (the
`wings') oriented at an angle to the active, or
high-surface-brightness, lobes (Leahy $\&$ Williams \cite{leahy}). Both sets of lobes often pass
symmetrically through the center of the host galaxy, 
giving the galaxy an X-shaped morphology as seen
on radio maps.  Several authors have attempted to explain the unusual
structure in X-shaped sources. Rees (\cite{rees}) suggested that the X-shaped morphology may result from a precession of the jet caused by misalignment between the angular momenta of the central supermassive black hole (SMBH) and the gas accreted onto it.
Subsequently, four different scenarios were proposed
for the formation of this peculiar radio morphology: {\em 1)}~backflow from
the active lobes into the wings (Leahy \& Williams \cite{leahy};
Capetti et al. \cite{capetti}); {\em 2)}~slow conical precession of the jet
axis (Parma, Ekers \& Fanti \cite{parma}; Mack et al. \cite{mack});
{\em 3)}~reorientation of the jet axis during which flow continues; and {\em 4)}~reorientation of the jet axis, but the jet turned off or at greatly
reduced power during the change of direction (Dennett-Thorpe et
al. \cite{dennett}). Merritt \& Ekers (\cite{merritt}) suggested that
the reorientation of a black hole spin axis owing to a minor merger may
lead to a sudden flip in the direction of the associated jet. In this
case, X-shaped galaxies would be the sites of spin-flips associated
with recent coalescences of pairs of supermassive black holes 
(Gopal-Krishna et al. \cite{gopal}; Gergely $\&$ Biermann
\cite{gergely}).  The older wings would then represent relics of past
radio jets, while the active lobes would reflect the activity that ensued after
the black hole merger. Hence, synchrotron aging should lead to a
steeper spectral index in the wings or low-surface brightness features
than in the high-surface brightness active lobes. However, in some
X-shaped sources, the wings have a similar or even flatter
spectral index than the active lobes (Lal \& Rao \cite{Lal}). To
explain this discrepancy Lal \& Rao (\cite{Lal}) suggested that at
least some of the X-shaped sources may contain coalescing binary SMBH
systems with two pairs of jets associated with two unresolved
AGN. The activity of the secondary black holes in post-merger galaxies is
suggested to depend on the separation and mass ratio of the two
SMBH (Lobanov \cite{lobanov}), with equal mass mergers providing the
longest time during which both SMBH remain active, while minor mergers lead to rapid disruption of the accretion disks around the secondary black hole and to quench of its activity (the primary black hole in minor merger systems may however remain active for a much longer time compared with the activity duration in equal mass mergers, {\em cf.}, Dokuchaev \cite{dokuchaev})

Several studies have been carried out in the optical and X-ray band, with results in some cases favoring the hydrodynamic backflow model over the merger-induced rapid reorientation scenario ({\em e.g.}, Hodges-Kluck et al. \cite{hodges710}; Landt et al. \cite{landt2010}), and in other cases favoring the merger model as the best explanation for the observed winged morphology (Hodges-Kluck et al. \cite{hodges717}b). 
A more detailed review of the different scenarios proposed to explain the origin of X-shaped radio galaxies is presented by Gopal-Krishna et al. (\cite{gopal2010}).

In all schemes in which X-shaped
objects are the product of galactic mergers,
the properties of the nuclear region
will be affected by the past SMBH merger or the presence of a
secondary SMBH. This should
then be reflected by statistically higher masses of the central black
holes in galaxies hosting the X-shaped radio. If the X-shaped morphology is indeed
caused by a profound event in the nuclear region, such as a merger of
two SMBH, then it may also be reflected in the starburst history of
the host galaxy (Blecha et al. \cite{blecha}). Investigations of the black hole masses and starburst histories can therefore address the
question of the physical nature of the X-shaped sources and their
difference from the ``canonical'' radio galaxies, if representative
samples of both radio types are studied and compared against each
other.

In this paper, we determine the black hole masses, luminosities, starburst histories,
and jet dynamic ages in a sample of
X-shaped galaxies and compare them with a control sample of radio-loud active nuclei with similar
redshifts and optical and radio luminosities. 

The target and control samples used for the study are introduced in
Sect.~2. Section~3 describes the data analysis and methods employed for determination of
the black hole masses and dynamic ages of the radio emission.
Results of these measurements are presented in
Sect.~4. A discussion of the results and their broader implication is
given in Sect. 5.

Throughout this paper, we assume a $\Lambda$--CDM
cosmology with parameters $H_{0}\ =\ 73\ km\ s^{-1}\ Mpc^{-1}$,
$\Omega_{\Lambda}\ = 0.73$ and $\Omega_{m}\ =\ 0.27$.

\section{The sample}

The sample of AGN analyzed here is drawn from a list of known
X-shaped sources and a list of 100 `winged' and X-shaped radio
source candidates (Cheung \cite{cheung}) that were morphologically identified
in the images from the VLA FIRST survey (Becker et al. \cite{becker}).

Of the 100 X-shaped candidates, 22 were found to have spectroscopic information
in the Sloan Digital Sky Survey (SDSS) data release (DR6; Adelman-McCarthy et al. \cite{adelman}),
and optical spectra for additional 27 objects were obtained by Cheung et
al. (\cite{cheung09}) with the 9.2 m Hobby-Eberly Telescope (HET) at
McDonald Observatory and the 6.5 m Multiple-Mirror Telescope (MMT) at
Mt. Hopkins Observatory.  

Cheung et al. (2009) have defined a subsample of 50 radio galaxies
with bona fide X-shaped radio morphology. Because this selection was based on visual inspection of the candidate objects and lacks quantifiable selection criteria, we refer to the original sample of Cheung (2007) for selecting appropriate candidates for our study. We select the X-shaped objects with optical spectra that show well-detected stellar absorption lines. Our final sample
comprises 18 X-shaped radio sources with SDSS spectra complemented by
five of the 27 X-shaped objects with spectra presented in Cheung et
al. (\cite{cheung09}), and six of the previously known X-shaped radio
sources: 3C192, 4C+32.25, 4C+48.29 and 1059+169 (Lal \& Rao
\cite{Lal07}), 3C223.1 (Lal \& Rao \cite{Lal}) and 4C+01.30 (Wang et
al. \cite{wang}) with spectra also available in the SDSS.

In order to evaluate the results obtained for the X-shaped radio galaxies,
we compile a control sample of 19 radio-loud sources from Marchesini et
al. (\cite{marchesini}) plus 6 Fanaroff-Riley type II (FR II) sources
from de Vries et al. (\cite{vries}), and 11 radio loud elliptical
galaxies from Gon{\'a}lez-Serrano $\&$ Carballo (\cite{gonzalez}) that have SDSS spectra, and cover the same ranges of redshift ($z<0.3$) and optical and radio luminosities as the objects in the target sample. The resulting common luminosity ranges for both samples are: log $\lambda L_{5100 \AA}\ \in$\ [43.0, 46.0] and log $\nu L_\mathrm{1.4GHz}\ \in$\ [39.0, 44.5].
According to the Kolmogorov-Smirnov test (KS-test) the two samples differ at $0.9\sigma$, which warrants making statistical comparisons between them.

\section{Data analysis}

The optical spectra of the objects in both samples have been used to
measure properties of H$\beta$ and [OIII] emission lines and a number of stellar
absorption lines. These measurements were applied to making kinematic
mass estimates of the central black holes. The absorption line fits
were also used to recover starburst histories in the host
galaxies and to estimate the epochs of the most recent starburst. Angular
sizes of active lobes and inactive wings (in X-shaped objects) were
measured to determine the dynamic ages of the radio emission
in the X-shaped objects and in the objects from the control sample.

\subsection{Stellar absorption lines}

The SDSS spectra collected for the objects in both samples contain
several significant stellar absorption features (such as Ca H$+$K
$\lambda\lambda$3969, 3934, the Mg I b $\lambda\lambda$5167, 5173,
5184 triplet, and the Ca II $\lambda\lambda$8498, 8542, 8662 triplet,
etc.) that can be matched against a combination of different synthetic
stellar template spectra, yielding an estimate of the stellar velocity
dispersion.

We use the stellar population synthesis code STARLIGHT
(Asari et al. \cite{asari07}; Cid Fernandes et al. \cite{cid04,cid05,cid07}; Mateus et al. \cite{mateus06})
to model the observed spectrum $O_{\lambda}$.
The best fit is obtained using
a linear combination of simple stellar populations (SSPs) from the stellar library of Bruzual \&
Charlot (\cite{bruzual}) and a power-law component representing the AGN continuum emission.
In our fitting, we apply the standard stellar library consisting of 150 SSPs, and complement the fit with up to six power-law components given by $F(\lambda) = 10^{20} (\lambda/4020)^{\beta}$, where $\beta$ = -0.5, -1, -1.5, -2, -2.5, -3. Each SSP spans across six metallicities, Z = 0.005, 0.02, 0.2, 0.4, 1, and 2.5 $Z_{\odot}$, with 25 different ages between 1 Myr and 18 Gyr. The Galactic extinction caused by the foreground dust screen is modeled and parametrized by the V-band extinction, A$_{V}$. We adopt the extinction law of Cardelli et al. (\cite{cardelli}).

The resulting model spectrum $M_{\lambda}$ (combining an SSP and a power-law continuum components) is
\begin{equation} \label{equation2}
M_{\lambda}(x, M_{\lambda_{0}}, A_{V}, v_{*},\sigma_{*})=M_{\lambda_{0}}[\sum_{j=1}^{N_{*}}x_{j}b_{j,\lambda}r_{\lambda}]\otimes G(v_{*},\sigma_{*})\, ,
\end{equation}
where $b_{j,\lambda}\equiv
L_{\lambda}^{SSP}(t_{j},Z_{j})/L_{\lambda_{0}}^{SSP}(t_{j},Z_{j})$ is
the $j^{\mathrm th}$ template normalized at $\lambda_{0}$, \textit{x} is
the population vector, $M_{\lambda_{0}}$ is the synthetic flux at the
normalization wavelength, $r_{\lambda} \equiv 10^{-0.4 (A_{\lambda}-A_{\lambda_{0}})}$ is the reddening term, and
\textit{$G(v_{*},\sigma_{*})$} is the line-of-sight stellar velocity
distribution, modeled as a Gaussian feature centered at velocity
\textit{$v_{*}$} and broadened by $\sigma_{*}$. The best fit is
reached by minimizing $\chi^{2}$,
\begin{equation} \label{equation3}
\chi^{2}(x, M_{\lambda_{0}}, A_{V}, v_{*},\sigma_{*})=\sum_{\lambda=1}^{N_{\lambda}}[(O_{\lambda}-M_{\lambda})w_{\lambda}]^{2}\, ,
\end{equation}
where the weighted spectrum \textit{$w_{\lambda}$} is defined as the
inverse of the noise in the observed spectra.  
A more detailed description of the STARLIGHT code and its applications can be found in Asari et al. (\cite{asari07}); Cid Fernandes et al. (\cite{cid04,cid05,cid07}); Mateus et al. (\cite{mateus06}); Leon-Tavares et al. (\cite{leon-tavares}).

\begin{figure}
 \centering
\includegraphics*[width=\columnwidth]{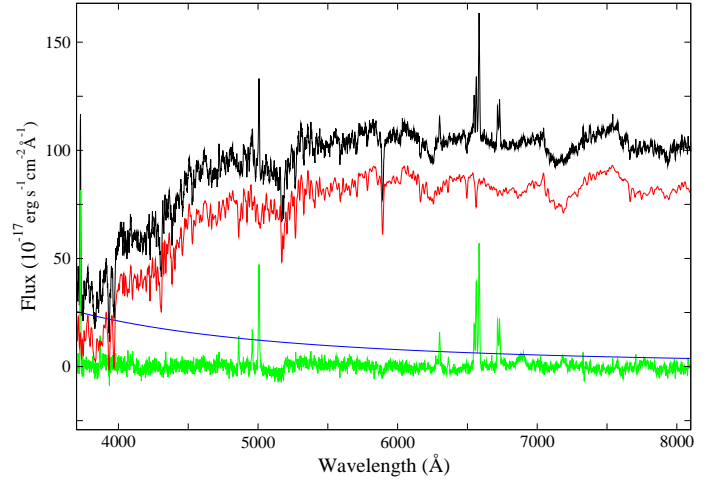}
  \caption{STARLIGHT fit for the spectrum of the X-shaped source J1424+2637. The plot shows the observed spectrum (top black line), 
the modeled spectrum (second red line, displaced from observed spectrum), the AGN continuum (blue line), and the residual spectrum (bottom green line).
Note that the residual spectrum contains narrow and (in some objects) broad emission lines. Properties of these lines are measured separately (see Sect. 3.2).}
\label{fig2}
\end{figure}

The fit for the X-shaped source J1424+2637 is shown as an example in Fig.~\ref{fig2}. The observed spectrum is shown in black, the host galaxy model in red, and the AGN power-law continuum in blue. The residual spectrum obtained after the subtraction of the stellar light and the continuum is shown in green.

In order to assess the fidelity of the STARLIGHT fit, we introduce a
quality factor $Q$ that combines the reduced $\chi^{2}$ parameter of the
modeled spectra, the velocity dispersion, $\sigma_{*}$, and its error,
$\delta_{*}$: $Q=(\chi^{2} \delta_{*}/\sigma_{*})^{-1/2}$.
Fits with $Q>10$ can be considered reliable (these are the fits with
reduced $\chi^2$ close to unity and fractional errors of the velocity
dispersion of $\le 3$\,\%).

For some objects with a strong continuum, STARLIGHT can fail to
synthesize the spectrum of the host galaxy, since the flux of the
lines of the host galaxy is much fainter than the AGN continuum
flux. In these cases, we apply the empirical correlation $\sigma_{*}$= FWHM
[OIII]/2.35 obtained by Nelson (\cite{nelson}) assuming that
$\sigma_{*} \approx \sigma_{gas}$. The term FWHM[OIII] describes the full-width-at-half-maximum of the narrow component of the [OIII] emission line
and is determined from a fit to the narrow lines of the
[OIII]$\lambda\lambda4959, 5007\AA$ doublet in the residual spectrum.

\subsection{Emission lines}

We determine the properties of the emission lines using the
residual spectra obtained after subtracting the host galaxy contribution and AGN
continuum from the STARLIGHT fits.  The H$\beta$ and [OIII] emission
lines in these spectra are modeled as a combination of the narrow and
broad components. We first fit three narrow Gaussian components for the
oxygen [OIII]4959$\AA$ and [OIII]5007$\AA$ lines and for the H$\beta$ line.
Three additional constraints are applied during the fitting
to reduce the number of free parameters and to provide robust
fits: {\em 1)}~the central wavelengths of the narrow line components
are set to their respective laboratory wavelengths; {\em 2)}~the FWHM
of the two [OIII] narrow components are required to have the same
value; {\em 3)}~the 1:3 ratio of amplitudes of the [OIII]4959$\AA$ and
[OIII]5007$\AA$ narrow components is fixed.

After adjusting the narrow components, broad Gaussian components are
added to the H$\beta$ line, and the joint fit is further adjusted
until the relative residuals are reduced to below 0.02-0.03.

\subsection{Starburst histories}

\begin{figure}
\centering 
\includegraphics[width=0.8\columnwidth]{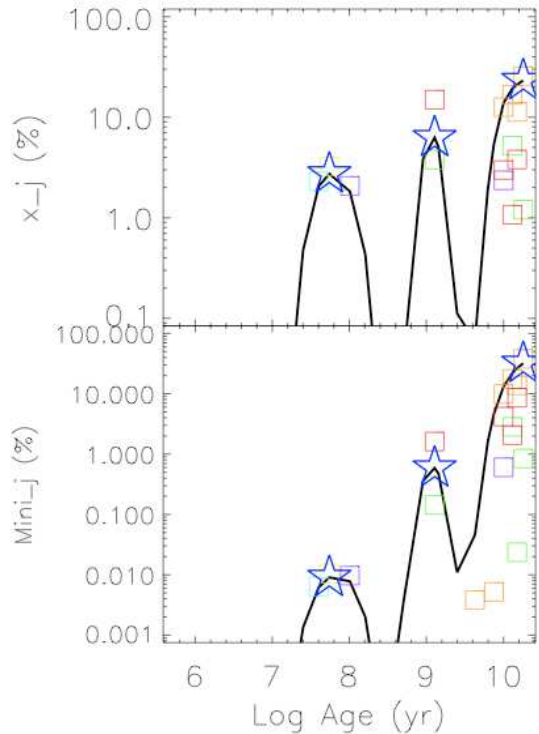}
  \caption{STARLIGHT fits for J1327$-$0203: the light fraction ($x_\mathrm{j}$) vs. age (top) and
  mass fraction $M\mathrm{ini}_\mathrm{j}$ vs. age (bottom) for each stellar population used. The squares correspond to stellar
  populations with different metallicities (in the color version: blue 0.005 Z$_{\odot}$, purple 0.020 Z$_{\odot}$, green 0.200 Z$_{\odot}$, yellow 0.400
  Z$_{\odot}$, orange 1.000 Z$_{\odot}$ and red 2.500 Z$_{\odot}$. The solid curve represents Gaussian smoothing applied to
  the mass and light fraction distributions of individual starbursts. The stars indicate the identified resulting starburst episodes.}
\label{fig11}
\end{figure}

The STARLIGHT model for the observed spectra also yields the light
fraction $x_\mathrm{j}$, mass fraction $M\mathrm{ini}_\mathrm{j}$, age
$\tau_\mathrm{j}$, and metallicity $Z_\mathrm{j}$, of the stellar
populations used in the fit. We use these parameters to derive
starburst histories and estimate the epochs of the most recent
starbursts in the studied galaxies. We apply Gaussian smoothing to
the individual starburst events (see Fig.~\ref{fig11}) and determine
 the epoch of the most recent starburst episode. 

\subsection{Optical continuum}
Most of the X-shaped radio sources and some objects from the control
sample are type II AGN (weakly beamed radio sources), for which STARLIGHT cannot provide reliable
estimates of the AGN continuum flux. 
We estimate the rest-frame continuum
flux at 5100 \AA\ from the SDSS photometry,
with the relation (Wu \& Liu \cite{wu})
\begin{equation}\label{equation7}
f_{5100\AA}\mathrm{[Jy]}=3631\ x\,10^{-0.4g}\ \left[\frac{4700}{5100(1+z)}\right]^{-(g-r)/2.5log(6231/4770)}\,
\end{equation}
where \textit{z} is the redshift and the \textit{g} and \textit{r} fiber
magnitudes are obtained from the SDSS and corrected for the Galactic
extinction $A_{V}$ (taken from Schlegel et al. \cite{schlegel}). It
should be noted that the flux obtained with this method also contains
a contribution from the host galaxy, but this does not
introduce a strong bias in our estimates.

To assess the spectral classification of the X-shaped and control host galaxies, we measure the Ca II break of their absoption optical spectrum.
This break is typically seen in the spectrum of elliptical galaxies and is described by a factor $C_\mathrm{Ca\,II} = (f_{+} - f_{-})/f_{+}$, where $f_{-}$ and $f_{+}$ are the fluxes in the rest frame wavelength regions 3750-3950 $\AA$ and 4050-4250 $\AA$, respectively (Landt et al. \cite{landt}). 
The Ca II break was used as an additional criterion to separate blazars from radio galaxies. Stocke et al. (\cite{stocke}) adopted a maximum value of $C_\mathrm{Ca\,II} = 0.25$ for BL Lacs to ensure the presence of a substantial non-thermal jet continuum in addition to the thermal spectrum of the elliptical host galaxy. This limit was later increased to $C_\mathrm{Ca\,II} <0.4$ by other authors (Marcha et al. \cite{marcha}; Plotkin et al. \cite{plotkin}).

Landt et al. (\cite{landt2010}) provide the Ca II break values of 16 of the X-shaped sources studied here. For the rest of the X-shaped sources and control objects, we measure the $C$ factor of the rest-frame optical spectra using the IRAF task guiapps.spectool.

\subsection{Black hole masses}

The measured stellar velocity dispersion, $\sigma_{*}$, can be
connected with the mass, $M_\mathrm{BH}$, of the central black hole
through an empirical relation (Gebhardt et al. \cite{gebhardt};
Tremaine et al. \cite{tremaine}):
\begin{equation} \label{equation1}
M_\mathrm{BH}=1.349\ x\,10^{8}\ M_{\odot}\ \left(\frac{\sigma_{*}}{200\ \mathrm{km\ s^{-1}}}\right)^{4.02\pm0.32}\
\end{equation}.
This relation is valid under the assumption that the kinematics of the
stars in the bulge of the host galaxy is dominated by the
gravitational potential of the central SMBH (Ferrarese \& Merritt
\cite{ferrarese}).

\subsection{Dynamic age of radio lobes}

The dynamic age, $t_\mathrm{a}$, of the high-surface-brightness (active)
radio lobes is obtained from their angular size, $\theta_\mathrm{a}$,
which is defined as the separation between the center of the radio source and
the most distant contour in the FIRST image. The dynamic age is then
given by $\theta_\mathrm{a}/v_\mathrm{a}$, where $v_\mathrm{a}$ is the
lobe advance speed (we adopt the commonly assumed $v_\mathrm{a}
\approx 0.1\,c$; {\em cf.}, Tingay et al. \cite{tingay}).  Assuming
that the fuelling of the low-surface-brightness lobes of the X-shaped
sources had stopped after the high-surface ones were activated, the
dynamic age of the passive lobes $t_{p}$ during their active stage can
be estimated as
\begin{equation}\label{wings}
 t_{p}= \frac{\theta_{p}-t_{a}v_{p}}{v_{a}}\,,
\end{equation}
where $\theta_{p}$ is the angular size of the low-surface brightness
lobes and $v_{p}$ is their expansion speed during the inactive
stage. In the absence of an observational estimate for $v_\mathrm{p}$, we
use $v_{p}= 0.01\,c$ in our calculation. It should be noted that
reducing $v_\mathrm{p}$ further only has a small effect on the ages
derived for the passive lobes.  The total dynamic age of the X-shaped
sources can then be obtained from the sum of the dynamic ages of the
active and passive lobes.

\section{Results}

The combined results from fitting the optical spectra, black hole mass
calculations, and age measurements for the radio lobes and most recent
starbursts are presented in Tables~2--3.

Tables 2 and 3 list (for the X-shaped objects and the control sample,
respectively) the object name based on J2000.0 coordinates (Col. 1), 
other common catalog names (Col. 2), stellar velocity dispersion (Col. 3), black hole mass
derived from $\sigma_{*}$ (Col. 4), optical luminosity of the AGN (Col. 5) and
of the host galaxy (Col. 6), radio luminosity (Col. 7), dynamic age of the radio lobes (Col. 8),
most recent starburst age (Col. 9), spectroscopic redshift (Col. 10), quality factor of the STARLIGHT fit (Col. 11),
the value of Ca II break factor (Col. 12), and references for the control sources (Table 3, Col. 13).  For the X-shaped
sources, the total (active $+$ passive lobe) age of the radio emission
is given in brackets in Col. 8.

The control sample objects 1217+023, 2349-014, 1004+130, 3C277.1 and 3C254 are
quasars with strong power-law continuum in their spectra.  Estimates of $\sigma_{*}$ for these objects are obtained from the measured width of the [OIII] line as described in Sect. 3.1.
The STARLIGHT results for two X-shaped sources (J0941-0143 and
J1348+4411) have provided unreliable $\sigma_{*}$ fits (with values values of $\sigma_{*}$ below the instrumental resolution of $70\,$km/s in the SDSS spectra,
and quality factors Q $<$10) and are excluded from further analysis.

\subsection{Luminosity matching}
\begin{figure}
 \centering \includegraphics[width=\columnwidth,
  clip=true]{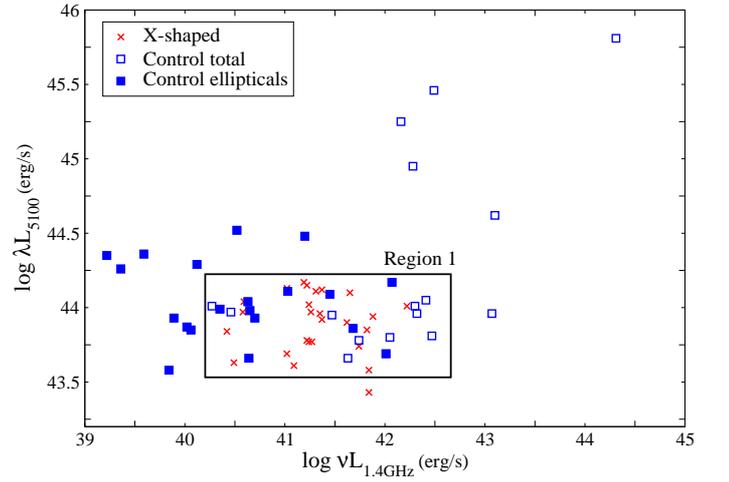} \caption{Optical continuum luminosity
  versus radio luminosity at 1.4 GHz for X-shaped (crosses) and control sources 
(squares). A small sub-region ("Region 1" marked by a rectangle inside the plot) is identified to provide a tighter luminosity match between the two samples.
In the further analysis, Region 1 is also subdivided into two equal bins in the radio luminosity to assess a possible luminosity dependence of the black hole mass estimates.}
\label{fig4}
\end{figure}

Luminosity matching between the target and control samples is illustrated in Fig. 3, where the continuum luminosities derived from the SDSS
magnitudes (Eq.~\ref{equation7}) are plotted against the radio luminosities at 1.4 GHz. The entire original radio-optical luminosity range, including all 
sources, is called Region 0 hereafter.  To provide a tighter match
between the samples (at the expense of reducing the sample
sizes), we define a smaller subregion or window (shown in
Fig.~\ref{fig4}) called Region 1. 
Region 1 ranges from log $\lambda L_{5100 \AA}\ \in$\ [43.5, 44.25] to log $\nu L_{1.4GHz}\ \in$\
[40.25, 42.5] and mainly excludes the control sources with the lowest radio
luminosities and highest optical luminosities because there are no X-shaped sources with such luminosities.  
The KS-test indicates that
the statistical difference between the two samples in Region 1 is 0.9$\sigma$ for the optical luminosity and 1.0$\sigma$ for the radio luminosity.

\subsection{Spectral classification}
We base our spectral classification on color-color separation and on the Ca II break measurements in the target and control samples.
All X-shaped sources analyzed have C$_{Ca\ II}$ $>$ 0.25, with 92$\%$ of the sources (23 out of 25) having C$_{Ca\ II}$ $\simeq$ 0.4. These values ensure that
the host galaxy of the X-shaped objects is elliptical and that it is dominated by the thermal spectrum, without any significant contribution of the non-thermal jet continuum.

For the control sample, however, the Ca II break can be determined for only 20 out of the 36 sources. Ninety percent of these sources (18 out of 20) have C$_{Ca\ II}$ $>$ 0.25, and 70$\%$ (14 out of 20) have C$_{Ca\ II}$ $\simeq$ 0.4. Only 56$\%$ of the control sources therefore have a thermal spectrum that dominates the host elliptical galaxy.

\begin{figure}
 \centering
 \includegraphics*[width=\columnwidth]{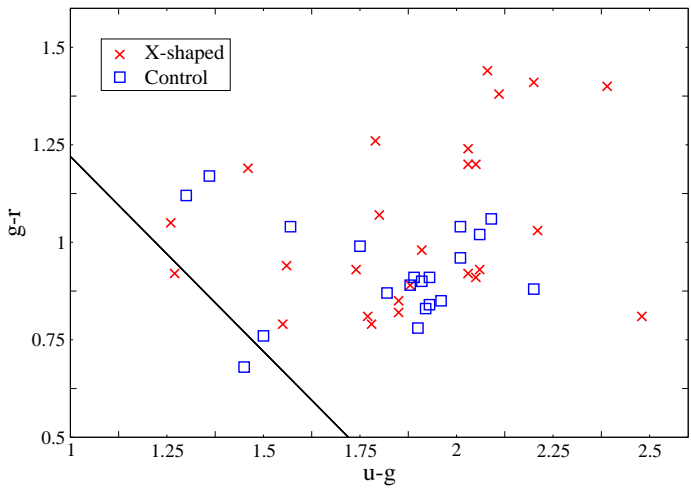}
  \caption{Color-color diagram ({\em g-r} colors versus {\em u-g} colors) for
X-shaped sources (cross) and control sample (square).  Black line:
$u-r= 2.22$ galaxy type separator from Strateva et
al. (\cite{strateva}). Sources situated above the line are classified as
ellipticals.}
\label{fig5}
\end{figure}

The spectral classification of the X-shaped and control sources can be further refined with color-color diagrams. We use the \textit{g}, \textit{r} and \textit{u} magnitudes obtained from SDSS to plot a \textit{g}-\textit{r} vs \textit{u}-\textit{g} color-color diagram (Fig.~\ref{fig5}) for
the two samples. We include in this plot only the sources with a spectrum dominated by the host galaxy (as determined from the Ca II break).
According to the distribution of galaxies in the {\em u-g}
vs {\em g-r} diagram from Strateva et al. (\cite{strateva}), all
galaxies lying above the $u-r = 2.22$ separator line are elliptical
systems. This is the case for all the X-shaped radio sources and for all the control sources for which
the SDSS magnitudes are available, with only one source from each of the samples lying slightly below the $u-r = 2.22$ separator line.
The X-shaped source J0049+0059  (with $u-g = 5.0$) is not plotted in Fig.~\ref{fig5}, but it has been classified as an elliptical galaxy as well.
This spectral classification agrees with the results obtained using the Ca II break feature and accordingly confirms that all X-shaped analyzed here are hosted by elliptical galaxies.
  
In order to better study the relation between the
galactic hosts of the X-shaped and control radio sources, we will also
consider a control subsample of elliptical galaxies from now on.
This subsample is defined according to the galactic type obtained by
the SDSS color separation described above and the Ca II break, and contains the 20 control sources for which the $C_\mathrm{Ca\,II}$ value could be determined.

\subsection{Black hole masses}
In Table 1 we show the mean black hole mass for each sample and subsamples 
and for each of the regions
described above (Regions 0 and 1).  The ratios of the mean black
hole mass defined as $r_\mathrm{xc} = \langle
M_\mathrm{BH,X-shaped}\rangle/\langle M_\mathrm{BH,control}\rangle$
are given as well.

\begin{table*}
\caption{Mean black hole masses and mass ratios}
\centering
\begin{tabular}{cc|ccccc}
\hline
\hline 
 Region & & & X-shaped & Control (all) & Control (ellipticals) \\
\hline
\multicolumn{6}{c}{} \\
 0 & & $\langle M_\mathrm{BH}\rangle [M_{\odot}]$ & $21.81^{+3.42}_{-2.96}\ \times\ 10^{7}$ & $14.78^{+2.35}_{-2.03}\ \times\ 10^{7}$ & $19.38^{+3.49}_{-2.95}\ \times\ 10^{7}$ \\
    & & & & & & \\
    &  & $r_\mathrm{xc}$ & & $1.48^{+0.24}_{-0.20}$ & $1.13^{+0.20}_{-0.17}$ \\
\hline  
    & & & & & & \\
  1 & & $\langle M_\mathrm{BH}\rangle [M_{\odot}]$ & $21.03^{+3.33}_{-2.87}\ \times\ 10^{7}$ & $10.87^{+2.34}_{-1.93}\ \times\ 10^{7}$ & $13.98^{+3.38}_{-2.72}\ \times\ 10^{7}$  \\
    & & & & & & \\
    & & $r_\mathrm{xc}$ & & $1.94^{+0.42}_{-0.34}$ & $1.50^{+0.36}_{-0.29}$ \\
\hline
    & & & & & & \\
  1 & Bin 1 & $\langle M_\mathrm{BH}\rangle [M_{\odot}]$ & $23.39^{+4.62}_{-3.86}\ \times\ 10^{7}$ & $15.05^{+4.23}_{-3.30}\ \times\ 10^{7}$ & $14.10^{+5.02}_{-3.70}\ \times\ 10^{7}$  \\
    & & & & & & \\
    & & $r_\mathrm{xc}$ & & $1.55^{+0.44}_{-0.34}$ & $1.66^{+0.59}_{-0.44}$ \\
    & & & & & & \\
  & Bin 2 & $\langle M_\mathrm{BH}\rangle [M_{\odot}]$ & $16.00^{+4.20}_{-3.33}\ \times\ 10^{7}$ & $8.75^{+2.71}_{-2.07}\ \times\ 10^{7}$ & $14.03^{+8.83}_{-5.42}\ \times\ 10^{7}$  \\
    & & & & & & \\
    & & $r_\mathrm{xc}$ & & $1.83^{+0.57}_{-0.43}$ & $1.14^{+0.72}_{-0.44}$ \\
\hline
\end{tabular}
\label{table1}
\end{table*}

A comparison of the black hole masses derived for the entire
samples (objects in Region 0) yields ratios of mean BH mass
ranging between 1.48 for X-shaped/control (all) to 1.13 for
X-shaped/control (ellipticals).

The mass ratios derived for objects in Region 1 (tighter common range of luminosities) are 
higher, with the mean BH mass of the X-shaped sample being nearly
twice the one derived for the whole control sample.   
In order to further reduce any potential effect of the source luminosity on the
derived black hole mass, we calculate mass ratios in two equal-size bins
in Region 1:
bin 1 ranges from log $\nu L_{1.4GHz}\ \in$\ [40.25,
41.4] and bin 2 from log $\nu L_{1.4GHz}\ \in$\ [41.5, 42.5]. 
One can see from the resulting ratios given in Table 1 that the
mass ratio does not show any evolution between the bins for the whole control sample.
The decrease of the mass ratio in bin 2 for the subsample of control ellipticals is owing to the small number of sources in this bin (3 sources).

\begin{figure}
 \centering
\includegraphics[width=\columnwidth, clip=true]{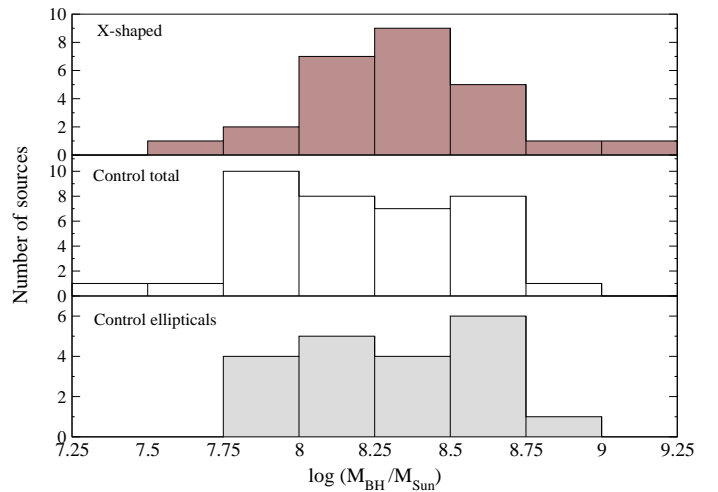}
  \caption{Histogram of the black hole mass in Region 0 for X-shaped sources (top), control all (middle), and control ellipticals (bottom).}
\label{fig6}
\end{figure}

\begin{figure}
 \centering
\includegraphics[width=\columnwidth, clip=true]{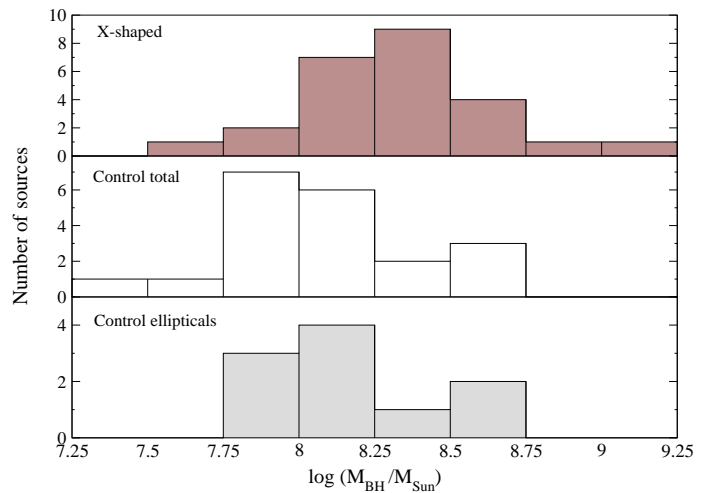}
  \caption{Histogram of the black hole mass in Region 1 for X-shaped sources (top), control all (middle), and control ellipticals (bottom).}
\label{fig7}
\end{figure}

The differences in the black hole masses are
further illustrated in Figs.~\ref{fig6}--\ref{fig7}, which show
histograms of the mass distributions for X-shaped and control samples
in Regions 0 and 1, respectively. In both regions the X-shaped objects
show a tendency to have higher black hole masses. 
This trend is more significant in the tighter Region 1, where 60$\%$ of the X-shaped sources have log M$_{BH}$ $>$ 8.25 M$_{\odot}$ 
while only 25$\%$ of the whole control sample does.
.

The KS-test applied to the BH masses of the X-shaped and control
objects indicates that the mass distributions are different at a
statistical significance of 1.6$\sigma$.  The differences are slightly
larger for objects in Region 1 (1.9$\sigma$). 

The statistically larger black hole mass of the X-shaped sample
implies that these objects are possibly located in galaxies that have
undergone strong major activity in the past, with either one
major merger event or multiple minor mergers.

\subsection{Starbursts}

The starburst and radio lobe ages can be used to see whether starburst
activity is different in the X-shaped objects and whether it can be
related to the production of radio emission in both the target and
control samples.

\begin{figure}[!h]
 \centering
  \includegraphics[width=\columnwidth, clip=true]{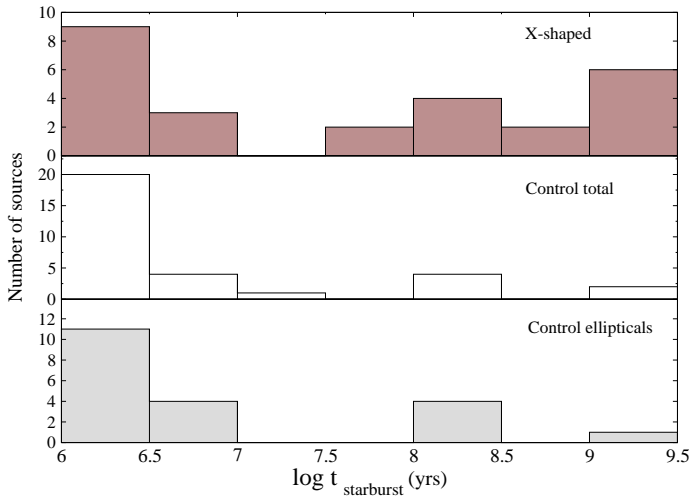}
  \caption{Histogram of the most recent starburst for X-shaped sources (top), control all (middle), and control ellipticals (bottom).}
\label{fig8}
\end{figure}

A histogram of the ages of the most recent starburst (Fig.~\ref{fig8})
in the objects from Region 0 shows that a sizeable fraction of objects
in both samples have relatively recent starbursts, with ages of
$10^{6.0}$--$10^{6.5}$ years, which are possibly related to the current
jet activity.  The distribution of starburst ages in the
X-shaped sample is much broader, with half of the objects
having starbursts older than $10^8$ years, while only 19$\%$ of all the
control sources and 25$\%$ of the control ellipticals exhibit these
starbursts. Because the largest dynamic age of the active lobes is $\sim10^{7}$ years 
(see Tables 2 and 3, Col. 8), any starburst activity significantly older than $10^{7}$ years is not likely to be
related to the active lobes. It can be speculated that these older starbursts are
related to galactic mergers themselves or to the putative coalescence
of the central black holes in post-merger galaxies. The latter
possibility can be viably tested by comparing the starburst ages to
the dynamic ages of the active and passive lobes in the X-shaped
objects.

The KS-test applied to the distributions
of the most recent starburst ages gives a statistical significance of
2.3$\sigma$ for the two samples being different. To account for a possible
dependence of this difference on the galactic type, we apply the
KS-test to the X-shaped sample and the subsample of control
ellipticals. The results of the test show that the starburst ages of
the X-shaped sources and the control ellipticals are still different
at a statistical significance of 2.1$\sigma$.

\begin{figure}[!h]
 \centering
  \includegraphics[width=\columnwidth, clip=true]{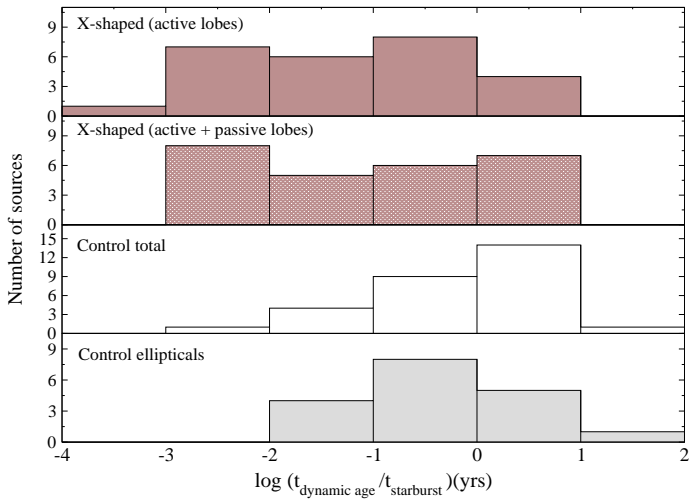}
  \caption{Logarithmic ratios of the dynamic age of the radio emission to the
age of the most recent starburst for Region 0. The top row presents the ages of active
lobes in X-shaped objects, while the second row shows the sum of the
ages of the active and passive lobes. The two bottom rows show
distributions of the age ratios in the control sample (third row) and
the subsample of control ellipticals (fourth row).}
\label{fig9}
\end{figure}

In Fig.~\ref{fig9}, histograms of the logarithm of the ratio of the
dynamic age and most recent starburst age are plotted for the X-shaped
sources and control sources of Region 0. 
The mean logarithmic ratios are $-1.29 \pm 0.23$ and $-0.14
\pm 0.18$ for the X-shaped objects and the control sample,
respectively.
The X-shaped sources tend to have 
starburst ages that are older than the dynamic ages of the radio lobes, 
while in both the control sample and the control
subsample of ellipticals these ages are comparable. The starburst activity in
X-shaped sources is therefore likely not related to the active
lobes. The KS-test shows that the ratio distributions are different at a statistical significance of 2.8$\sigma$. 
This difference in the starburst/dynamic age ratios may support the
scenario in which the active lobes of the X-shaped sources
are due to a possible reorientation caused by a black hole merger
(Merritt \& Ekers \cite{merritt}) that leaves the old low-surface-brightness lobes inactive.  
Assuming that the low-surface-brightness lobes became inactive when the high-surface 
ones were activated, the dynamic age of the passive lobes during their active stage can be
determined using Eq.~\ref{wings}.  The ratio of the total dynamic age
of the active plus passive lobes to the starburst age is plotted in 
Fig.~\ref{fig9} (second panel), and it indicates that the starburst age 
still remains older than the total dynamic age
of the lobes. This suggests that the starburst activity in
X-shaped sources had occured before the possible reorientation owing to a
black hole merger, and it may have been related to the galactic merger
itself.

Instead of analyzing all the X-shaped sources with optical spectra available, we can constrain the X-shaped sample to only the X-shaped radio galaxies included in the bona fide sample of Landt et al. (\cite{landt2010}). This implies taking out 10 of the 29 X-shaped sources included in our sample, which leads to a 50$\%$ increase in the statistical erros, but does not change the obtained results.
The statistical studies would be improved, on the other hand, by the addition of more X-shaped radio sources to the sample. This will be done with the availability of new optical spectra.

\section{Summary}
We determined and studied the black hole masses
and starburst histories of 29 X-shaped radio sources 
to attempt to find out the most likely physical scenario for the formation of their peculiar morphology.
We compared the black
hole masses and properties of the starburst activity derived for
this sample with those obtained for a control sample of 36
galaxies with common optical and radio luminosities.  The main results
of this study can be summarized as follows.

The Ca II break feature of the absorption spectrum and the optical colors obtained from the SDSS $u,g,r$ photometry data
indicate that all X-shaped radio sources studied in this paper
are hosted by elliptical galaxies. 
In the hierarchical galaxy-formation model, today's galaxies are the 
product of frequent galaxy merging, triggering starburst and active 
galactic nuclei activity and possibly forming supermassive binary black hole systems (Frenk et al. \cite{frenk}; Hutchings $\&$ Campbell \cite{campbell}; Sanders et al. \cite{sanders}; Hopkins et al. \cite{hopkins}). 
There appears to be supporting evidence that elliptical galaxies have formed from mergers in a hierarchical way
(Toomre $\&$ Toomre \cite{toomre}; Efstathiou $\&$ Silk \cite{efstathiou}). 
Therefore the trend for X-shaped radio sources to reside in elliptical galaxies suggests that their
X-shaped morphology could be related to the galaxy merger scenario.

The mean black hole mass of the objects in the X-shaped sample is found to be
statistically higher than in the control
sample. This result holds for the entire range of the radio and
optical luminosities as well as for smaller subranges and for exact
galactic type-matching between the two samples.

The statistically higher black hole masses obtained for the X-shaped
objects suggest that their peculiar radio morphology can indeed result
from a reorientation of the jet axis owing to black hole coalescence or
from the presence of two active central engines in the nuclear region.

A comparison of the starburst ages and dynamic ages of the radio
lobes reveals that the most recent starburst activity in X-shaped
sources is older than in the control sample and that X-shaped radio
sources had their most recent starburst before their active lobes were
formed. These results lend further support to the scenario in which
the high-surface-brightness lobes might have become active due to
reorientation caused by a black hole coalescence, while the peak of starburst activity could be connected to the galactic merger itself.

Another scenario proposed to explain the X-shaped morphology in X-shaped radio galaxies employs the backflow emission of the active lobes into the wings. This model is based on the presence of pressure gradients in the confining medium of the jet and does not seem to be able to explain the results we obtained.
Although our findings argue in favor of the merger scenario, they do not rule out entirely the backflow model. 
It should be further noted that it is quite possible that some X-shaped sources may indeed have undergone a merger, while in others the peculiar shape might be caused by a backflow of the active jets.

\section*{Acknowledgments}

The authors are grateful to the anonymous referee for insightful comments and suggestions.
The authors thank C.C. Cheung for providing optical
spectroscopy data and M. Karouzos for his valuable assistance.
M. Mezcua was supported for this research through a stipend from the
International Max Planck Research School (IMPRS) for Radio and
Infrared Astronomy at the Universities of Bonn and Cologne.
This work was supported by the CONACYT research grant 54480 (M\'exico).
The STARLIGHT project is supported by the Brazilian agencies CNPq, CAPES and
FAPESP and by the France-Brazil CAPES/Cofecub program.


\begin{landscape}
\begin{table}
\small
\begin{minipage}[t]{\columnwidth}
\caption{\label{table2} X-shaped objects}
\centering
\begin{tabular}{cccccccccccc}
\hline
\hline
\multicolumn{2}{c}{Name} & $\sigma_\star$ & $\log\,M_\mathrm{BH}$ & $\log\,\lambda L_\mathrm{opt}^{agn}$ & $\log\,\lambda L_\mathrm{opt}^{gal}$ & $\log\,\nu L_\mathrm{rad}$ & $\log\,t_\mathrm{a}\,\,(\log\,t_\mathrm{p})$ & $\log\,t_\mathrm{sb}$ &  $z$ & $Q$ & $C_\mathrm{Ca\,II}$\\
J2000        & Other      & [km/s]          & [$M_\odot$]   & [erg/s]        & [erg/s]& [erg/s]& [yr]      & [yr] &          &       &      \\ 
(1)          & (2)        & (3)             & (4)           & (5)            & (6)   & (7)   & (8)         & (9)  & (10)     & (11)  & (12) \\ \hline
J0001$-$0033	&		&	277.32	$\pm$	4.55	&	8.70	$\pm$	0.03	&	$^{-b}$			&	43.77	&	41.27	&	6.37	\quad	(6.60)	&	6.50	&	0.247	&	83.94	&	0.43$^f$	\\
J0049$+$0059	&		&	282.23	$\pm$	9.64	&	8.73	$\pm$	0.06	&	$^{-b}$			&	43.43	&	41.84	&	6.68	\quad	(6.83)	&	9.20	&	0.304	&	31.07	&		\\
J0808$+$2409	&	3C192	&	209.94	$\pm$	7.97	&	8.21	$\pm$	0.07	&	$^{-b}$			&	43.85	&	41.82	&	6.52	\quad	(6.71)	&	6.70	&	0.060	&	41.46	&	0.36	\\
J0813$+$4347	&		&	251.25	$\pm$	5.79	&	8.53	$\pm$	0.04	&	$^{-b}$			&	44.11	&	41.31	&	6.29	\quad	(6.49)	&	9.10	&	0.128	&	44.69	&	0.45$^f$	\\
J0831$+$3219	&	4C$+$32.25	&	194.80	$\pm$	6.21	&	8.08	$\pm$	0.06	&	42.42	$\pm$	0.14	&	43.97	&	41.26	&	6.69	\quad	(6.85)	&	6.50	&	0.051	&	59.44	&	0.44$^f$	\\
J0838$+$3253	&		&	241.60	$\pm$	11.43	&	8.46	$\pm$	0.08	&	$^{-b}$			&	43.92	&	41.37	&	6.59	\quad	(6.76)	&	8.50	&	0.213	&	28.77	&	0.42	\\
J0859$-$0433	&		&	250.37	$\pm$	12.90	&	8.52	$\pm$	0.09	&	42.33	$\pm$	2.65	&	$^{-c}$	&	42.13	&	6.08	\quad	(6.65)	&	8.00	&	0.356	&	24.55	&		\\
J0924$+$4233	&		&	231.16	$\pm$	15.72	&	8.38	$\pm$	0.12	&	$^{-b}$		-	&	43.58	&	41.84	&	6.54	\quad	(6.77)	&	9.00	&	0.227	&	15.96	&	0.42$^f$	\\
J0941$-$0143	&		&	154.92	$\pm$	20.61	&	7.68	$\pm$	0.23	&	42.48	$\pm$	5.15	&	43.19	&	42.84	&	6.14	\quad	(6.64)	&	7.00	&	0.384	&	4.92$^e$	&		\\
J0941$+$3944	&	3C223.1	&	196.90	$\pm$	9.14	&	8.10	$\pm$	0.08	&	42.30	$\pm$	0.87	&	43.94	&	41.88	&	6.44	\quad	(6.78)	&	9.11	&	0.107	&	35.30	&	0.38$^f$	\\
J1005$+$1154	&		&	272.88	$\pm$	6.46	&	8.67	$\pm$	0.04	&	$^{-b}$		-	&	44.12	&	41.37	&	6.43	\quad	(6.67)	&	9.10	&	0.166	&	66.08	&	0.48$^f$	\\
J1020$+$4831	&	4C$+$48.29	&	194.30	$\pm$	7.94	&	8.08	$\pm$	0.07	&	42.28	$\pm$	0.19	&	43.78	&	41.22	&	6.72	\quad	(7.02)	&	8.46	&	0.053	&	38.82	&	0.44$^f$	\\
J1040$+$5056	&		&	211.57	$\pm$	15.88	&	8.23	$\pm$	0.13	&	$^{-b}$		-	&	43.77	&	41.24	&	6.44	\quad	(6.65)	&	9.20	&	0.154	&	20.74	&	0.42	\\
J1043$+$3131	&		&	195.27	$\pm$	8.24	&	8.09	$\pm$	0.07	&	41.15	$\pm$	1.15	&	43.63	&	40.49	&	5.80	\quad	(6.09)	&	6.80	&	0.036	&	32.66	&	0.41	\\
J1101$+$1640	&	1059$+$169	&	220.98	$\pm$	6.42	&	8.30	$\pm$	0.05	&	41.01	$\pm$	5.18	&	43.69	&	41.02	&	6.62	\quad	(6.98)	&	6.00	&	0.069	&	60.03	&	0.45$^f$	\\
J1111$+$4050	&		&	254.75	$\pm$	4.96	&	8.55	$\pm$	0.03	&	$^{-b}$		-	&	44.17	&	41.19	&	6.16	\quad	(6.43)	&	6.50	&	0.074	&	74.31	&	0.43	\\
J1130$+$0058	&	4C$+$01.30	&	151.86	$\pm$	6.78	&	7.65	$\pm$	0.08	&	43.12	$\pm$	0.15	&	44.10	&	41.65	&	6.24	\quad	(6.61)	&	9.40	&	0.132	&	24.88	&	0.26$^f$	\\
J1140$+$1057	&		&	196.37	$\pm$	7.80	&	8.10	$\pm$	0.07	&	40.51	$\pm$	20.11	&	43.61	&	41.09	&	6.28	\quad	(6.51)	&	6.00	&	0.081	&	40.99	&	0.41$^f$	\\
J1207$+$3352	&		&	181.28	$\pm$	3.79	&	7.96	$\pm$	0.04	&	41.85	$\pm$	0.84	&	44.13	&	41.02	&	6.19	\quad	(6.44)	&	6.00	&	0.079	&	51.36	&	0.28$^f$	\\
J1210$-$0341	&		&	221.29	$\pm$	9.90	&	8.30	$\pm$	0.08	&	42.14	$\pm$	2.94	&	43.90	&	41.62	&	5.97	\quad	(6.56)	&	8.70	&	0.178	&	33.61	&	0.40$^f$	\\
J1210$+$1121	&		&	223.51	$\pm$	9.41	&	8.32	$\pm$	0.07	&	$^{-b}$		-	&	43.96	&	41.35	&	6.66	\quad	(6.78)	&	7.60	&	0.196	&	33.55	&	0.42	\\
J1327$-$0203	&		&	237.30	$\pm$	9.27	&	8.43	$\pm$	0.07	&	$^{-b}$		-	&	44.01	&	42.22	&	6.39	\quad	(6.65)	&	7.70	&	0.183	&	38.98	&	0.48$^f$	\\
J1330$-$0206	&		&	219.35	$\pm$	7.72	&	8.29	$\pm$	0.06	&	$^{-b}$		-	&	44.04	&	40.59	&	6.35	\quad	(6.61)	&	7.00	&	0.087	&	42.40	&	0.40	\\
J1339$-$0016	&		&	323.77	$\pm$	4.55	&	8.97	$\pm$	0.02	&	$^{-b}$		-	&	44.15	&	41.22	&	6.55	\quad	(6.81)	&	8.50	&	0.145	&	88.09	&	0.44	\\
J1348$+$4411	&		&	108.42	$\pm$	27.71	&	7.05	$\pm$	0.45	&	41.67	$\pm$	10.43	&	43.28	&	41.63	&	5.87	\quad	(6.53)	&	6.90	&	0.267	&	2.95$^e$	&		\\
J1424$+$2637	&		&	174.99	$\pm$	6.53	&	7.90	$\pm$	0.07	&	40.48	$\pm$	6.37	&	43.97	&	40.58	&	6.13	\quad	(6.37)	&	6.50	&	0.037	&	45.36	&	0.40	\\
J1444$+$4147	&		&	226.06	$\pm$	9.35	&	8.34	$\pm$	0.07	&	$^{-b}$		-	&	43.74	&	41.74	&	6.60	\quad	(6.80)	&	8.50	&	0.188	&	35.89	&	0.44$^f$	\\
J1455$+$3237	&		&	224.17	$\pm$	5.52	&	8.33	$\pm$	0.04	&	41.57	$\pm$	1.78	&	43.84	&	40.42	&	6.10	\quad	(6.31)	&	6.50	&	0.084	&	68.39	&	0.39	\\
J1614$+$2817	&		&	344.13	$\pm$	6.21	&	9.08	$\pm$	0.03	&	41.32	$\pm$	2.65	&	44.02	&	41.24	&	5.59	\quad	(6.22)	&	6.00	&	0.108	&	25.85	&	0.42$^f$	\\
\hline
\end{tabular}
\end{minipage}
\smallskip\newline {\bf Column designation:}~(1) -- object name based on J2000.0 coordinates; 
(2) -- other common catalog names; (3) -- stellar velocity dispersion obtained from STARLIGHT; 
(4) -- black hole mass obtained from $\sigma_{*}$; (5) -- 5100 \AA\ continuum
luminosity from STARLIGHT; (6) -- 5100 \AA\ continuum luminosity from
SDSS photometry; (7) -- 1.4 GHz radio luminosity; (8) -- dynamic age of the
active (active+passive) lobes; (9) -- age of the most recent starburst; (10) -- spectroscopic
redshift from SDSS; (11) -- quality factor; (12) value of Ca II break factor . {\bf Notes:}~$a$ -- velocity dispersion
obtained using the correlation $\sigma_{*}$=FWHM [OIII]/2.35.$; b$ -- STARLIGHT could 
not fit the continuum luminosity; $c$ -- no SDSS photometry available; $d$ -- no STARLIGHT fit;
$ e$ -- low fidelity of the STARLIGHT fit; $f$ -- values of Ca II break from Landt et al. (\cite{landt2010})
\end{table}
\end{landscape}

\begin{landscape}
\begin{table}
\tiny
\begin{minipage}[t]{\columnwidth}
\caption{\label{table3} Control sample}
\centering
\begin{tabular}{ccccccccccccc}
\hline
\hline
\multicolumn{2}{c}{Name} & $\sigma_\star$ & $\log\,M_\mathrm{BH}$ & $\log\,\lambda L_\mathrm{opt}^{agn}$ & $\log\,\lambda L_\mathrm{opt}^{gal}$ & $\log\,\nu L_\mathrm{rad}$ & $\log\,t_\mathrm{a}$ & $\log\,t_\mathrm{sb}$ & $z$ & $Q$ & $C_\mathrm{Ca\,II}$ & Ref.\\
J2000 & Other & [km/s] & [$M_\odot$] & [erg/s] & [erg/s] & [erg/s] & [yr] & [yr] &  &  &  &  \\ 
(1)          & (2)        & (3)             & (4)           & (5)            & (6)   & (7)   & (8)  & (9)  & (10) & (11)  & (12)  & (13) \\
 \hline
J0758$+$3747	&	0755+37	&	251.55	$\pm$	4.63	&	8.53	$\pm$	0.03	&	41.58	$\pm$	0.09	&	44.48	&	41.20	&	6.16	&	6.00	&	0.043	&	85.50	&	0.36	&	M04	\\
J0803$+$2440	&	B2 0800+24	&	186.04	$\pm$	0.13	&	8.00	$\pm$	0.06	&	$^{-b}$			&	43.85	&	40.06	&	6.42	&	6.50	&	0.043	&	55.56	&	0.42	&	G00	\\
J0821$+$4702	&	3C197.1	&	144.35	$\pm$	9.53	&	7.56	$\pm$	0.12	&	42.46	$\pm$	0.05	&	43.80	&	42.05	&	6.11	&	6.00	&	0.128	&	17.33	&	-	&	M04	\\
J0822$+$0557	&	3C198	&	164.35	$\pm$	5.76	&	7.79	$\pm$	0.06	&	42.46	$\pm$	0.02	&	43.66	&	41.63	&	6.43	&	6.00	&	0.081	&	51.31	&	-	&	M04	\\
J0846$+$3126	&	B2 0843+31	&	183.02	$\pm$	0.14	&	7.98	$\pm$	0.07	&	$^{-b}$			&	43.58	&	39.84	&	6.98	&	7.00	&	0.067	&	43.13	&	0.41	&	G00	\\
J0921$+$4538	&	3C219	&	188.98	$\pm$	8.49	&	8.03	$\pm$	0.08	&	42.52	$\pm$	0.07	&	43.96	&	43.07	&	6.94	&	6.00	&	0.174	&	21.14	&	-	&	M04	\\
J0925$+$3627	&	4C36.14	&	256.37	$\pm$	0.35	&	8.56	$\pm$	0.04	&	$^{-b}$			&	44.26	&	39.36	&	6.60	&	7.00	&	0.112	&	68.51	&	0.43	&	G00	\\
J0939$+$3553	&	3C223	&	181.42	$\pm$	8.26	&	7.96	$\pm$	0.08	&	42.45	$\pm$	0.06	&	43.81	&	42.47	&	7.06	&	9.40	&	0.137	&	24.50	&	-	&	M04	\\
J0941$+$5751	&	J0941+5751	&	189.32	$\pm$	7.83	&	8.03	$\pm$	0.07	&	$^{-b}$			&	43.97	&	40.46	&	6.43	&	6.00	&	0.159	&	29.56	&	-	&	V06	\\
J0947$+$0725	&	3C227	&	112.51	$\pm$	3.35	&	7.13	$\pm$	0.05	&	43.63	$\pm$	0.00	&	43.95	&	41.47	&	6.78	&	6.00	&	0.086	&	24.43	&	-	&	M04	\\
J1006$+$2554	&	B2 1003+26	&	240.82	$\pm$	0.40	&	8.45	$\pm$	0.06	&	$^{-b}$			&	44.52	&	40.52	&	7.17	&	9.11	&	0.117	&	38.18	&	0.44	&	G00	\\
J1006$+$3454	&	3C236	&	247.41	$\pm$	7.26	&	8.5	$\pm$	0.05	&	$^{-b}$			&	44.17	&	42.07	&	7.74	&	6.50	&	0.101	&	54.96	&	0.33	&	M04	\\
J1007$+$1248	&	1004+130	&	167.82$^{a}$	$\pm$	1.05	&	7.77	$\pm$	0.01	&	$^{-b}$			&	45.46	&	42.49	&	6.89	&	$^{-d}$	&	0.241	&	$^{-d}$	&	-	&	M04	\\
J1031$+$5225	&	J1031+5225	&	204.85	$\pm$	11.06	&	8.17	$\pm$	0.09	&	42.34	$\pm$	0.10	&	43.69	&	42.01	&	6.42	&	6.00	&	0.166	&	22.09	&	0.28	&	V06	\\
J1040$+$2957	&	4C30.19	&	161.97	$\pm$	0.06	&	7.76	$\pm$	0.04	&	$^{-b}$			&	44.11	&	41.03	&	$^{-c}$	&	8.01	&	0.091	&	61.46	&	0.18	&	G00	\\
J1055$+$5202	&	J1055+5202	&	162.90	$\pm$	12.40	&	7.77	$\pm$	0.13	&	42.61	$\pm$	0.11	&	43.78	&	41.74	&	6.46	&	6.00	&	0.187	&	12.12	&	-	&	V06	\\
J1105$+$3009	&	B2 1102+30A	&	262.26	$\pm$	0.35	&	8.60	$\pm$	0.04	&	41.14	$\pm$	6.59	&	44.35	&	39.22	&	6.35	&	6.50	&	0.072	&	74.22	&	0.43	&	G00	\\
J1114$+$4037	&	3C254	&	234.20$^{a}$	$\pm$	3.29	&	8.35	$\pm$	0.03	&	$^{-b}$			&	45.81	&	44.31	&	1.55	&	$^{-d}$	&	0.736	&	$^{-d}$	&	-	&	M04	\\
J1154$+$0238	&	J1154+0238	&	169.59	$\pm$	11.91	&	7.84	$\pm$	0.12	&	$^{-b}$			&	43.66	&	40.64	&	6.46	&	6.70	&	0.211	&	18.78	&	0.31	&	V06	\\
J1219$+$0549	&	3C270	&	269.46	$\pm$	3.72	&	8.65	$\pm$	0.02	&	$^{-b}$			&	44.04	&	40.63	&	6.12	&	7.00	&	0.007	&	84.46	&	0.41	&	M04	\\
J1220$+$0203   	&	1217+023	&	174.94$^{a}$	$\pm$	1.72	&	7,84	$\pm$	0.02	&	$^{-b}$			&	45.25	&	42.16	&	6.84	&	$^{-d}$	&	0.240	&	$^{-d}$	&	-	&	M04	\\
J1252$+$5634	&	3C277.1	&	235.90$^{a}$	$\pm$	0.57	&	8.37	$\pm$	0.00	&	$^{-b}$			&	44.62	&	43.10	&	5.04	&	$^{-d}$	&	0.320	&	$^{-d}$	&	-	&	M04	\\
J1259$+$2757	&	NGC4874	&	251.47	$\pm$	0.32	&	8.53	$\pm$	0.04	&	$^{-b}$			&	44.36	&	39.59	&	6.30	&	6.00	&	0.024	&	57.54	&	0.42	&	G00	\\
J1319$+$2938	&	4C29.47	&	198.73	$\pm$	0.17	&	8.12	$\pm$	0.06	&	$^{-b}$			&	44.09	&	41.45	&	6.35	&	8.01	&	0.073	&	53.97	&	0.39	&	G00	\\
J1321$+$4235  	&	3C285	&	162.58	$\pm$	9.46	&	7.77	$\pm$	0.10	&	$^{-b}$			&	43.86	&	41.68	&	6.6	&	6.00	&	0.079	&	26.37	&	0.24	&	M04	\\
J1332$+$0200	&	3C287.1	&	246.02	$\pm$	7.86	&	8.49	$\pm$	0.06	&	43.24	$\pm$	0.04	&	44.01	&	40.27	&	6.82	&	6.00	&	0.216	&	32.12	&	-	&	M04	\\
J1341$+$5344	&	J1341+5344	&	225.84	$\pm$	7.07	&	8.34	$\pm$	0.05	&	$^{-b}$			&	43.87	&	40.02	&	6.55	&	6.00	&	0.141	&	42.80	&	0.31	&	V06	\\
J1350$+$2816  	&	B2 1347+28	&	209.58	$\pm$	0.23	&	8.21	$\pm$	0.06	&	$^{-b}$			&	43.98	&	40.65	&	6.34	&	8.01	&	0.072	&	46.69	&	0.41	&	G00	\\
J1430$+$5201	&	3C303	&	167.41	$\pm$	9.77	&	7.82	$\pm$	0.10	&	43.22	$\pm$	0.01	&	43.96	&	42.32	&	6.28	&	7.20	&	0.141	&	16.86	&	-	&	M04	\\
J1512$+$0203	&	J1512+0203	&	194.50	$\pm$	10.45	&	8.08	$\pm$	0.09	&	42.79	$\pm$	0.09	&	44.01	&	42.30	&	6.52	&	6.00	&	0.219	&	25.33	&	-	&	V06	\\
J1529$+$3042	&	1527+30	&	322.89	$\pm$	7.49	&	8.97	$\pm$	0.01	&	$^{-b}$			&	44.29	&	40.12	&	6.38	&	8.20	&	0.114	&	64.60	&	0.46	&	G00	\\
J1559$+$2556	&	B2 1557+26	&	223.51	$\pm$	0.24	&	8.32	$\pm$	0.05	&	41.09	$\pm$	15.68	&	43.93	&	39.89	&	$^{-c}$	&	6.00	&	0.045	&	54.67	&	0.4	&	G00	\\
J1611$+$3103	&	B2 1609+31	&	199.34	$\pm$	0.23	&	8.12	$\pm$	0.08	&	$^{-b}$			&	43.93	&	40.70	&	5.74	&	6.00	&	0.095	&	31.15	&	0.42	&	G00	\\
J1615$+$2726	&	1613+27	&	221.13	$\pm$	6.13	&	8.31	$\pm$	0.05	&	41.11	$\pm$	0.40	&	43.99	&	40.35	&	5.89	&	6.00	&	0.065	&	58.17	&	0.42	&	M04	\\
J1617$+$3222	&	3C332	&	172.37	$\pm$	3.80	&	8.64	$\pm$	0.01	&	40.90	$\pm$	4.45	&	44.05	&	42.41	&	6.62	&	6.00	&	0.151	&	33.13	&	-	&	M04	\\
J2351$-$0109	&	2349-014	&	201.97$^{a}$	$\pm$	2.12	&	8.72	$\pm$	0.01	&	$^{-b}$			&	44.95	&	42.28	&	6.06	&	$^{-d}$	&	0.174	&	$^{-d}$	&	-	&	M04	\\
\hline
\end{tabular}
\end{minipage}
\smallskip\newline
 {\bf Column designation:}~(1) -- object name based on J2000.0 coordinates; 
(2) -- other common catalog names; (3) ; -- stellar velocity dispersion obtained from STARLIGHT; 
(4) -- black hole mass obtained from $\sigma_{*}$; (5) -- 5100 \AA\ continuum
luminosity from STARLIGHT; (6) -- 5100 \AA\ continuum luminosity from
SDSS photometry; (7) -- 1.4 GHz radio luminosity; (8) -- dynamic age of the
active lobes; (9) -- age of the most recent starburst; (10) -- spectroscopic
redshift from SDSS; (11) -- quality factor; (12) value of Ca II break factor; (13) -- Original references. {\bf Notes:}~$a$ -- velocity dispersion
obtained using the correlation $\sigma_{*}$=FWHM [OIII]/2.35.$; b$ -- STARLIGHT could 
not fit the continuum luminosity; $c$ -- not available; $d$ -- no STARLIGHT fit.
REFERENCES.-- (G00) Gon{\'a}lez-Serrano $\&$ Carballo (\cite{gonzalez}); (M04) Marchesini et al. (\cite{marchesini});
(V06) de Vries et al. (\cite{vries}).
\normalsize
\end{table}
\end{landscape}

\end{document}